\documentclass[article,preprint,groupedaddress]{revtex4}
\usepackage{epsfig}
\usepackage{graphicx}

\begin{document}
\title{On the Quantization of a Multi-Horizon Black Hole}
\author{Shahar Hod}
\address{The Ruppin Academic Center, Emeq Hefer 40250, Israel}
\address{ }
\address{The Hadassah Institute, Jerusalem 91010, Israel}
\date{\today}

\begin{abstract}

\ \ \ We consider the quasinormal spectrum of a charged scalar field
in the (charged) Reissner-Nordstr\"om spacetime, which has two
horizons. The spectrum is characterized by two distinct families of
asymptotic resonances. We suggest and demonstrate that according to
Bohr's correspondence principle and in agreement with the
Bekenstein-Mukhanov quantization scheme, one of these resonances
corresponds to a fundamental change of $\Delta A=4\hbar\ln2$ in the
surface area of the black-hole outer horizon. The second asymptotic
resonance is associated with a fundamental change of $\Delta
A_{tot}=4\hbar\ln3$ in the total area of the black hole (in the sum
of the surface areas of the inner and outer horizons), in accordance
with a suggestion of M\"akel\"a and Repo.

\end{abstract}
\bigskip
\maketitle

%]

It is generally believed that the surface area of black holes would
have a discrete character in a quantum theory of gravity. Based on
the remarkable observation that the minimal increase in black-hole
surface area has a universal character \cite{Beken1,Beken2},
Bekenstein suggested an area spectrum with a uniform spacing

\begin{equation}\label{Eq1}
\Delta A=\gamma \hbar \  ,
\end{equation}
where $\gamma$ is a dimensionless constant \cite{Rev}. (We use units
in which $G=c=k_B=1$ henceforth).

Years later, Mukhanov and Bekenstein \cite{Muk,BekMuk,Beken3} have
combined the black-hole area-entropy relation $S_{BH}=A/4\hbar$ and
the Boltzmann-Einstein formula in statistical physics in order to
fix the value of the unknown coefficient $\gamma$. They have found
that $\Delta A$ should be of the form

\begin{equation}\label{Eq2}
\Delta A=4\hbar \ln{k} \  ,
\end{equation}
where $k$ is some natural number.

Nevertheless, we were still left with the unknown value of the
integer $k$. In order to gain more information, one may apply Bohr's
correspondence principle to the characteristic resonances of the
black hole, its (discrete) quasinormal mode (QNM) spectrum
\cite{Hod2}. Black holes have an infinite number of (complex)
quasinormal frequencies, describing oscillations with decreasing
relaxation times \cite{Nollert1,Leaver,Nollert2,Andersson}. Based on
Bohr's correspondence principle, it was argued that the asymptotic
Schwarzschild resonances are given by \cite{Hod2}

\begin{equation}\label{Eq3}
\omega = \pm T^{s}_{BH}\ln3 -i 2\pi T^{s}_{BH} (n+{1\over2})\  ,
\end{equation}
where $T^{s}_{BH}=1/8\pi M$ is the Bekenstein-Hawking temperature of the Schwarzschild black hole.
An analytical proof of this equality was later given in \cite{Motl}.

The emission of a quantum of frequency $\omega$ results in a change
$\Delta M=\hbar \omega_R$ in the black-hole mass. Assuming that
$\omega$ corresponds to the asymptotically damped limit
\cite{Note1}, Eq. (\ref{Eq3}), and using the first-law of black-hole
thermodynamics $\Delta M={1 \over 4}T^{s}_{BH} \Delta A$, this
implies a fundamental change in the black-hole surface area

\begin{equation}\label{Eq3c}
\Delta A =4\hbar \ln3 \  .
\end{equation}
Thus, the correspondence principle, as applied to the black-hole
resonances, provides the missing link, and gives evidence in favor
of the value $k=3$.

The possible correspondence between the black-hole classical
resonances and the quantum properties of its surface area has
triggered a flurry of research attempting to calculate the
asymptotic ringing frequencies of various types of black holes (for
a detailed list of references see, e.g., \cite{HodKesh}). Recently,
we have determined the asymptotic resonances of a charged scalar
field in the (charged) Reissner-Nordstr\"om (RN) spacetime
\cite{Hodc}.

Remarkably, it turns out that the charged quasinormal spectrum of
the multi-horizon RN black hole has a unique feature: in contrast
with the single-horizon Schwarzschild black hole which has a single
asymptotic resonance, the multi-horizon RN black hole is
characterized by {\it two} distinct families of asymptotic QNM
resonances \cite{Hodc}:

\begin{equation}\label{Eq5}
\omega = \pm T_{BH}\ln2 +{{eQ} \over {r_+}} -i 2\pi T_{BH} (n+{1\over2})\  ,
\end{equation}
where $r_{\pm}=M \pm (M^2-Q^2)^{1/2}$ are the black hole (event and
inner) horizons, $T_{BH}=(r_{+}-r_{-})/4\pi r^2_{+}$ is the
Bekenstein-Hawking temperature of the black hole, $e$ is the charge
coupling constant of the field [$e$ stands for $e/\hbar$, and has
the dimensions of (length)$^{-1}$], and

\begin{equation}\label{Eq6}
\omega = \mp T^{s}_{BH}\ln3 +{{eQ} \over {r^s_+}} -i 2\pi T^{s}_{BH} (n+{1\over2})\  ,
\end{equation}
where $r^s_+=2M$ is the Schwarzschild radius \cite{Notec}. The
upper/lower signs correspond to positive/negative values of $eQ$,
respectively.

It has been demonstrated (see \cite{Hodc} for details) that the
fundamental resonance, Eq. (\ref{Eq5}) corresponds to a change of

\begin{equation}\label{Eq8}
\Delta A =4\hbar \ln2\  ,
\end{equation}
in the outer black-hole surface area, in remarkable agreement with
the Bekenstein-Mukhanov general prediction, Eq. (\ref{Eq2}).

The physical significance of the second branch, Eq. (\ref{Eq6}), was
less clear at the time \cite{Hodc}. Here we demonstrate that it can
be associated to an intriguing proposal of M\"akel\"a and Repo
\cite{Mak}. It has been argued \cite{Mak} that in the case of a
multi-horizon black hole, it is the {\it total} area (of both the
inner and outer horizons) which should be quantized in equal steps.
For the RN black hole one finds that the total surface area (of both
horizons) is given by $A_{tot}=4\pi(4M^2-2Q^2)$, which implies

\begin{equation}\label{Eq9}
\Delta A_{tot}=4\pi(8M \Delta M -4Q\Delta Q) \  .
\end{equation}

The emission of a quantum of frequency $\omega$ and an electric
charge $e$ results in a change $\Delta M=\hbar \omega_R$ in the
black-hole mass, and a change $\Delta Q=e$ in its charge.
Substituting the fundamental black-hole resonance, Eq. (\ref{Eq6})
into the relation (\ref{Eq9}), one finds

\begin{equation}\label{Eq10}
\Delta A_{tot} =4\hbar \ln3\  .
\end{equation}
It should be emphasized that although the asymptotic resonance, Eq.
(\ref{Eq6}) depends on the various parameters of both the black hole
and the charged field, the corresponding area spacing, Eq.
(\ref{Eq10}), is universal in the sense that it is independent of
the black-hole parameters, $M$ and $Q$, and also independent of the
field parameters, its charge coupling $e$ and its angular index $l$.
We also note that this fundamental area unit is actually the {\it
same} one found for the single-horizon Schwarzschild black hole
\cite{Hod2}, see Eq. (\ref{Eq3c}).

In summary, we have considered the QNM spectrum of a charged scalar
field in the multi-horizon, charged RN spacetime. The asymptotic
black-hole resonances are characterized by two distinct branches. We
have shown that according to Bohr's correspondence principle, the
first asymptotic resonance corresponds to a fundamental black-hole
area unit $\Delta A=4\hbar\ln2$, in accord with the
Bekenstein-Mukhanov quantization scheme.

The second resonance corresponds to a fundamental change in the
total black-hole area, $\Delta A_{tot}=4\hbar\ln3$. The universality
of this area spacing (i.e., its independence of the black-hole
parameters) implies that the total area of a multi-horizon black
hole may be quantized in equal steps, in accordance with the
proposal of M\"akel\"a and Repo \cite{Mak}.

We would like to end with a conjecture: the fact that the
multi-horizon RN black hole has two distinct families of resonances
suggests that the corresponding classical black hole may split into
two distinct families when quantized \cite{Note2}: one in which the
outer surface area of the black hole is quantized in equal steps (of
$4\hbar\ln2$), and the other in which the total area (of both
horizons) is quantized in equal steps (of $4\hbar\ln3$). The later
seems to be a natural extension of the Schwarzschild solution, since
both have the same area spacing, $4\hbar\ln3$.

\bigskip
\noindent
{\bf ACKNOWLEDGMENTS}
\bigskip

This research is supported by the Meltzer science Foundation. I
thank Uri Keshet for numerous discussions, as well as for a
continuing stimulating collaboration.

%\newpage


\begin{thebibliography}{99}
\bibitem{Notec} These QNM resonances are valid for charged black holes that
satisfy the condition $eQ \gtrsim r_+/r_-$ [this condition also
reads $Q/M \gtrsim (\hbar/\alpha A)^{1/6}$, where $\alpha$ is the
fine structure constant]. For semi-classical black holes
(characterized by $A \gg \hbar$) this condition is satisfied by even
slightly charged black holes.
\end{thebibliography}
\end{document}